\documentstyle[12pt,titlepage]{article}
\input epsfig.sty

\font\grande=cmr9.5 scaled \magstep4
\font\medio=cmr9.5 scaled \magstep2
\outer\def\beginsection#1\par{\medbreak\bigskip
      \message{#1}\leftline{\bf#1}\nobreak\medskip
\vskip-\parskip
      \noindent}

\def\laq{\raise 0.4ex\hbox{$<$}\kern -0.8em\lower 0.62
ex\hbox{$\sim$}}
\def\gaq{\raise 0.4ex\hbox{$>$}\kern -0.7em\lower 0.62
ex\hbox{$\sim$}}

\begin{document}
\bibliographystyle {unsrt}

\titlepage

\begin{flushright}
CERN-PH-TH/2009-261
\end{flushright}

\vspace{15mm}
\begin{center}
{\grande Spectator stresses and CMB observables}\\
\vspace{15mm}
 Massimo Giovannini 
 \footnote{Electronic address: massimo.giovannini@cern.ch} \\
\vspace{0.5cm}
{{\sl Department of Physics, Theory Division, CERN, 1211 Geneva 23, Switzerland }}\\
\vspace{1cm}
{{\sl INFN, Section of Milan-Bicocca, 20126 Milan, Italy}}
\vspace*{2cm}

\end{center}

\vskip 1.5cm
\centerline{\medio  Abstract}
The large-scale curvature perturbations induced by spectator anisotropic stresses are analyzed across the matter-radiation transition.  It is assumed that the anisotropic stress is associated with a plasma component whose energy density  is subdominant both today and prior to photon decoupling. The enforcement of the momentum constraint and the interplay with the neutrino anisotropic stress determine the regular initial conditions of the Einstein-Boltzmann hierarchy. The Cosmic Microwave Background observables 
have shapes and phases which differ both from the ones of the conventional adiabatic mode as well as from their 
non-adiabatic counterparts.   
\noindent

\vspace{5mm}

\vfill
\newpage
Supplementary anisotropic stresses may arise before matter-radiation equality for diverse physical reasons. For instance, on a phenomenological ground, one is often led to consider 
a putative anisotropic stress associated with the dark-matter component 
\cite{refdm}. Along a similar line it is also legitimate to consider the possibility 
that the dark-energy component is endowed with an appropriate anisotropic stress 
\cite{koi1}. The two possibilities mentioned so far are characterized by a common feature: in both cases the anisotropic stress is attributed to species which are dominant (or just slightly subdominant) today. This is the case for the dark-matter and for the dark-energy components in the $\Lambda$CDM paradigm whose 
basic parameters will be taken to coincide, for the purposes of the present analysis, with the ones determined by analyzing the WMAP 5yr data 
alone\footnote{For sake of concreteness the values of the cosmological parameters used 
in the present paper will be taken to coincide with the parameters arising from the 5yr best fit to the WMAP data alone, i.e., using 
standard notations,  
$(\Omega_{\mathrm{b}0},\, \Omega_{\mathrm{c}0}, \,\Omega_{\mathrm{\Lambda}}, \,h_{0}, \,n_{\mathrm{s}},\,\epsilon_{\mathrm{re}})= (0.0441,\, 0.214,\, 0.742,\, 0.719,\, 0.963,\,0.087)$ where $\epsilon_{\mathrm{re}}$ denotes the optical depth to reionization and $n_{\mathrm{s}}$ the 
spectral index of (adiabatic) curvature perturbations.}
\cite{WMAP51,WMAP52}.
There exist however physical situations in which the anisotropic stress does not belong to species which are dominant today. Examples in this direction are numerous and include, for instance, spectator scalar fields evolving (without being dominant) during the various stages of the life of the Universe or even large-scale magnetic fields (see, for instance, \cite{mg1} and references therein). There is an important caveat to the latter statement: unlike large-scale magnetic fields, 
gravitating anisotropic stresses do not couple to the charged species of the plasma. The gravitational and electromagnetic effects associated with 
weakly coupled plasmas can be used to understand pre-decoupling magnetism in the presence of the relativistic fluctuations 
of the geometry and this approach leads to the calculation of what are customarily called magnetized Cosmic Microwave Background (CMB) observables\footnote{The inclusion 
of an anisotropic stress associated with a large-scale magnetic field 
 is a necessary requirement for the analysis of magnetized CMB anisotropies. It is though not sufficient since the coupling 
 to charged particles does lead both to gravitational and electromagnetic effects whose specific account is crucial to deduce both 
 the temperature anisotropies and 
 the polarization power spectra. Background magnetic fields also affect the photon-electron scattering in a computable way \cite{mg1a}.}. The purpose of the present paper is to investigate, in general terms, the interplay between a supplementary anisotropic stress and the other sources of anisotropic stress 
which are present in the system prior to photon decoupling. If the source of large-scale  inhomogeneity resides in a spectator 
anisotropic stress it will be interesting to compute the dynamical evolution of the relativistic fluctuations of the geometry as well as the temperature and polarization  correlations. It will be argued that the  ``stressed" initial conditions for the CMB anisotropies 
constitute an intermediate case between the purely adiabatic and the purely non-adiabatic initial conditions (see, e.g., \cite{hannu1}). 

The time evolution of the system\footnote{Without loss of generality the space-time geometry will be taken to be 
conformally flat implying that the background geometry can be written as $\overline{g}_{\mu\nu} =a^2(\tau) \eta_{\mu\nu}$ where 
$\eta_{\mu\nu}\equiv \mathrm{diag}(1,\, -1,\, -1,\,-1)$ is the Minkowski metric with signature mostly minus.}  will be parametrized in terms of the scale factor $a(\tau)$ normalized at equality, i.e. $\alpha = a/a_{\mathrm{eq}}$; within this parametrization 
the (total) barotropic index and the (total) sound speed can be written, respectively, as
\begin{eqnarray}
&&w_{\mathrm{t}}(\alpha) = \frac{p_{\mathrm{t}}}{\rho_{\mathrm{t}}}= \frac{1}{3 (\alpha +1)}, \qquad 
c_{\mathrm{st}}^2(\alpha) = w_{\mathrm{t}} - \frac{1}{3} \frac{\partial \ln{(w_{\mathrm{t}} +1)}}{\partial \ln{\alpha}}= \frac{4}{3(3\alpha +4)}.
\label{eq1}
\end{eqnarray}
The exact solution of the Friedmann-Lema\^itre equations in a spatially flat 
background geometry implies that $\alpha(x) = (x^2 + 2 x)$ where $x = \tau/\tau_{1}$
and $\tau$ is the conformal time coordinate. The evolution equations for the fluctuations 
of the geometry and of the sources shall be written in Fourier space and in terms 
of the rescaled wavenumber $\kappa = k \tau_{1}$; if $\kappa <1$ the given wavelength 
exceeds the Hubble radius right before the equality time. In the opposite 
case (i.e. $\kappa >1$) the given wavelength is smaller than the Hubble radius at the same time $\tau_{1} = (\sqrt{2} -1) \tau_{\mathrm{eq}}$ which is roughly half of the 
equality time when non-relativistic matter and radiation contribute equally to the expansion.  In the synchronous coordinate system the perturbed entry of the geometry reads\footnote{The conventions employed in 
Refs. \cite{MB1,MB2} differ from the ones employed here (and match with the ones of \cite{mg1} and references therein). The differences result, in a nutshell, from a different signature of the metric, from a different naming of the perturbed degrees of freedom and from a slightly different convention in the Rayleigh expansion of the brightness perturbations. 
Bearing in mind Eq. (\ref{eq2}) potential ambiguities are avoided.}
\begin{equation}
\delta_{\mathrm{s}} g_{ij}(\kappa, \alpha) = a_{\mathrm{eq}}^2 \,\alpha^2(\tau) \biggl[ \hat{\kappa}_{i} 
\hat{\kappa}_{j} h(\kappa,\alpha) + 6 \xi(\kappa,\alpha) \biggl(\hat{\kappa}_{i} \hat{\kappa}_{j} - \frac{\delta_{ij}}{3} \biggr)\biggr],
\label{eq2}
\end{equation}
where $h(\kappa,\alpha)$ and $\xi(\kappa,\alpha)$ describe the scalar fluctuations 
of the metric and enter also the contravariant components of the energy-momentum 
tensor\footnote{In what follows the arguments of the various functions will be omitted
and it will be assume, when not otherwise stated, that they do depend both on $\alpha$ and $\kappa$.}, i.e. 
\begin{eqnarray}
&& \delta_{\mathrm{s}} T^{00} = \frac{\delta_{\mathrm{s}} \rho_{\mathrm{t}}}{a_{\mathrm{eq}}^2 \alpha^2}, 
\qquad \delta_{\mathrm{s}} T^{0i} = \frac{1}{a_{\mathrm{eq}}^2\alpha^2} (p_{\mathrm{t}} + \rho_{\mathrm{t}}) v_{\mathrm{t}}^{i},
\nonumber\\
&& \delta_{\mathrm{s}} T^{ij} = \frac{1}{a_{\mathrm{eq}}^2\alpha^2}\biggl\{ \delta_{\mathrm{s}} p_{\mathrm{t}} \delta^{ij} + 2 p_{\mathrm{t}} \biggl[ - \xi \delta^{ij} + 
\frac{\hat{\kappa}^{i} \hat{\kappa}^{j}}{2} (h+ 6 \xi) \biggr] - \Pi^{ij} \biggr\}.
\label{eq3}
\end{eqnarray}
The fluctuations of the energy density and of the total anisotropic stress are\footnote{Using standard 
notations $\omega_{\mathrm{c}0} = h_{0}^2 \Omega_{\mathrm{c}0}$, 
$\omega_{\mathrm{b}0} = h_{0}^2 \Omega_{\mathrm{b}0}$ and $\omega_{\mathrm{M}0} = h_{0}^2 \Omega_{\mathrm{M}0}$.}
\begin{eqnarray}
&& \delta_{\mathrm{s}} \rho_{\mathrm{t}}= \rho_{\mathrm{t}} \biggl[
\Omega_{\mathrm{R}}\biggl( R_{\gamma} \delta_{\gamma} + R_{\nu} \delta_{\nu}\biggr)
+ \Omega_{\mathrm{M}} \biggl( \frac{\omega_{\mathrm{c}0}}{\omega_{\mathrm{M}0}}
\delta_{\mathrm{c}} + \frac{\omega_{\mathrm{b}0}}{\omega_{\mathrm{M}0}}
\delta_{\mathrm{b}}\biggr)\biggr],
\label{eq4}\\
&& \kappa_{i} \kappa_{j} \Pi^{ij}= (p_{\mathrm{t}} + \rho_{\mathrm{t}}) \kappa^2 \sigma_{\mathrm{t}} = \kappa^2 [(p_{\nu} + \rho_{\nu})  \sigma_{\nu} + \sum_{a} (p_{a} + \rho_{a}) \sigma_{a} + (p_{\gamma} + \rho_{\gamma}) \sigma_{x}],
\label{eq5}
\end{eqnarray}
where $\Omega_{\mathrm{R}}(\alpha) = 1/(\alpha + 1)$ and $\Omega_{\mathrm{M}}(\alpha) = \alpha/(\alpha+1)$.
In Eq. (\ref{eq4}) the density contrasts for neutrinos, photons, CDM 
particle and baryons have been introduced (i.e., respectively, $\delta_{\nu}$, $\delta_{\gamma}$, $\delta_{\mathrm{c}}$ and $\delta_{\mathrm{b}}$); in Eq. (\ref{eq5})
$\sigma_{\nu}(\kappa,\alpha)$ accounts for the neutrino anisotropic stress while $\sigma_{x}(\kappa,\alpha)$ denotes 
the anisotropic stress of a putative component which is subdominant before 
equality; $\sigma_{x}(\kappa,\alpha)$ will be referred 
to the photons since photons (together with neutrinos) constitute the dominant component  of the plasma before equality (i.e. for $\alpha < 1$).  In Eq. (\ref{eq5}) a sum over other potential components has been added: these components 
could account for the possible stresses arising either in the dark matter or in 
the dark energy sectors (see, e.g., \cite{refdm}) but will not be explicitly considered hereunder. 
In the $\alpha$-parametrization the Hamiltonian and the momentum constraints read, respectively, 
\begin{eqnarray}
&& \frac{\partial h}{\partial\alpha} = \frac{\kappa^2 \alpha}{2 (\alpha+ 1)} \xi - \frac{3}{\alpha}
\biggl[\Omega_{\mathrm{R}}\biggl(R_{\nu} \delta_{\nu} + R_{\gamma} \delta_{\gamma}\biggr) + \Omega_{\mathrm{M}} 
\biggl( \frac{\omega_{\mathrm{c}0}}{\omega_{\mathrm{M}0}}
\delta_{\mathrm{c}} + \frac{\omega_{\mathrm{B}0}}{\omega_{\mathrm{M}0}}
\delta_{\mathrm{b}}\biggr)\biggr],
\label{eq6}\\
&& \kappa^2 \alpha^2 \frac{\partial \xi}{\partial \alpha} = - \frac{4}{\sqrt{\alpha+1}} \biggl\{ R_{\nu} \theta_{\nu} 
+ R_{\gamma} [ 1 + R_{\mathrm{b}}(\alpha)] \theta_{\gamma\mathrm{b}} + \frac{3}{4} \frac{\omega_{\mathrm{c}0}}{\omega_{\mathrm{M}0}} \alpha \theta_{\mathrm{c}}\biggr\}.
\label{eq7}
\end{eqnarray}
To write Eq. (\ref{eq7}) the following identity has been employed:
\begin{equation}
(p_{\mathrm{t}} + \rho_{\mathrm{t}}) \theta_{\mathrm{t}} = \frac{4}{3} \rho_{\mathrm{t}}\biggl\{ \Omega_{\mathrm{R}} [ R_{\nu} \theta_{\nu}  + R_{\gamma} (1 + R_{\mathrm{b}}) \theta_{\gamma\mathrm{b}}] + \frac{3}{4} \Omega_{\mathrm{M}} \frac{\omega_{\mathrm{c}0}}{\omega_{\mathrm{M}0}} \theta_{\mathrm{c}}\biggr\},
\label{eq7a}
\end{equation}
where $\theta_{\mathrm{t}} = i \vec{\kappa}\cdot \vec{v}_{\mathrm{t}}$; moreover, in Eqs. (\ref{eq7})--(\ref{eq7a}) $R_{\nu}$, $R_{\gamma}=(1-R_{\nu})$ and $R_{\mathrm{b}}(\alpha)$ denote, respectively, the neutrino fraction, the photon fraction and the baryon-to-photon ratio
\begin{equation}
R_{\nu} = \frac{3 \times(7/8)\times (4/11)^{4/3}}{1 + 3 \times(7/8)\times (4/11)^{4/3}} =0.4052,\qquad R_{\mathrm{b}}(\alpha) 
= \frac{3}{4 R_{\gamma}} \biggl(\frac{\omega_{\mathrm{b}0}}{\omega_{\mathrm{M}0}}\biggr) \, \alpha \simeq 0.215 \, \alpha,
\label{eq7b}
\end{equation}
where $3$ counts the massless neutrino families, $(7/8)$ stems from the Fermi-Dirac statistics and $(4/11)^{4/3}$ is related 
to the different kinetic temperature of neutrinos; the numerical expression of $R_{\mathrm{b}}(\alpha)$ follows from the 
choice of parameters listed before in this paper.  
The remaining two equations stemming from the perturbed Einstein equations 
can be written as:
\begin{eqnarray}
&& \frac{\partial^2 h}{\partial \alpha^2} + \frac{5 \alpha + 4}{2 \alpha (\alpha +1)} \frac{\partial h}{\partial\alpha} - \frac{\kappa^2 \xi}{2 (\alpha + 1)}= \frac{3}{\alpha^2 (\alpha+1)}\biggl[ R_{\gamma} \delta_{\gamma} + R_{\nu} \delta_{\nu} + 3 w_{x} R_{\gamma} \Omega_{x}\biggr],
\label{eq8}\\
&& \frac{\partial^2 {\mathcal Q}}{\partial \alpha^2} + \frac{5 \alpha + 4 }{2 \alpha(\alpha+1)} \frac{\partial {\mathcal Q}}{\partial \alpha}= 
\frac{\kappa^2 \xi}{2 (\alpha + 1)} + \frac{12}{\alpha^2 (\alpha +1)} (R_{\nu} \sigma_{\nu} 
+ R_{\gamma} \sigma_{x}).
\label{eq9}
\end{eqnarray}
The evolution equations of the neutrinos obey, 
\begin{eqnarray}
&& \frac{\partial \delta_{\nu}}{\partial \alpha} = - \frac{2 \theta_{\nu}}{3 \sqrt{\alpha +1}} + \frac{2}{3} \frac{\partial h}{\partial \alpha}, 
\qquad \frac{\partial \theta_{\nu}}{\partial \alpha} = \frac{\kappa^2}{8 \sqrt{\alpha +1}} \delta_{\nu} 
- \frac{\kappa^2 }{2 \sqrt{\alpha +1}} \sigma_{\nu},
\label{eq10}\\
&& \frac{\partial {\sigma}_{\nu}}{\partial \alpha} = \frac{2 \theta_{\nu}}{15 \sqrt{\alpha + 1}} - \frac{2}{15} 
\frac{\partial {\mathcal Q}}{\partial \alpha}- \frac{3}{20} \frac{\kappa {\mathcal F}_{\nu\,3}}{\sqrt{\alpha +1}} ,
\label{eq10a}\\
&& \frac{\partial {\mathcal F}_{\nu\,\ell}}{\partial \alpha} = \frac{\kappa}{2 ( 2 \ell + 1)\sqrt{\alpha + 1}} 
[\ell {\mathcal F}_{\nu\, (\ell -1)} - (\ell +1) {\mathcal F}_{\nu\, (\ell +1)}],
\label{eq11}
\end{eqnarray}
where ${\mathcal Q}$ denotes the combination $(h + 6 \xi)$ and ${\mathcal F}_{\nu\ell}$ denotes the $\ell$th multipole 
of the perturbed phase-space distribution of the neutrinos. The evolution equations of the dark-matter sector obey instead 
\begin{equation}
\frac{\partial \delta_{\mathrm{c}}}{\partial \alpha}= - \frac{\theta_{\mathrm{c}}}{2 \sqrt{\alpha +1}} + \frac{1}{2} \frac{\partial h}{\partial \alpha}, \qquad \frac{\partial \theta_{\mathrm{c}}}{\partial \alpha} + \frac{\theta_{\mathrm{c}}}{\alpha} =0
\label{eq12}
\end{equation}
The system of photons an baryons can be treated, within an excellent approximation, 
to lowest order in the tight-coupling expansion; in the latter approximation 
the quadrupole of the photons vanishes and the resulting equation is given by
\begin{eqnarray}
&&\frac{\partial \theta_{\gamma\mathrm{b}}}{\partial \alpha} + \frac{R_{\mathrm{b}}\,\theta_{\gamma\mathrm{b}}}{\alpha ( R_{\mathrm{b}} +1)}  = \frac{\kappa^2\,\delta_{\gamma}}{8 \sqrt{\alpha +1} (R_{\mathrm{b}} +1)} ,
\label{eq13}\\
&& \frac{\partial \delta_{\gamma}}{\partial \alpha} = - \frac{2}{3} \frac{\theta_{\gamma\mathrm{b}}}{\sqrt{\alpha +1}} + \frac{2}{3} \frac{\partial h}{\partial \alpha},\qquad 
\frac{\partial \delta_{\mathrm{b}}}{\partial \alpha} = - \frac{\theta_{\gamma\mathrm{b}}}{2 \sqrt{\alpha +1}} + \frac{1}{2} \frac{\partial h}{\partial \alpha},
\label{eq14}
\end{eqnarray}
where $\theta_{\gamma\mathrm{b}}= \theta_{\gamma} \simeq
\theta_{\gamma\mathrm{b}}$ represents the common value of the photon and baryon velocities.  
Covariant conservation of the total energy-momentum tensor implies that $\sigma_{x}(\kappa,\alpha) = 3 \, w_{x} \, \Omega_{x}(\kappa,\alpha)/4$ as well as $\Omega_{x}(\kappa,\alpha) = [4/(3 w_{x})] {\mathcal B}(\kappa) (\alpha/\alpha_{\mathrm{i}})^{1-3w_{x}}$. Needless to say that various manipulations 
can be used, in different physical limits, to obtain approximate expressions. For instance, by dropping the velocity contribution 
in Eq. (\ref{eq10a}) as well as ${\mathcal F}_{\nu 3}$, an approximate evolution equation for $\sigma_{\nu}$ can be derived from Eq. 
(\ref{eq9}), i.e. 
\begin{equation}
\frac{\partial^2 \sigma_{\nu}}{\partial\alpha^2} + \frac{5 \alpha + 4}{2 \alpha(\alpha+1)} \frac{\partial \sigma_{\nu}}{\partial \alpha} + \frac{\kappa^2 \xi}{15 (\alpha + 1)} + \frac{8}{5\alpha^2(\alpha+1)} (R_{\nu} \sigma_{\nu} + R_{\gamma}\sigma_{x}) \simeq 0.
\label{eq15}
\end{equation}
While Eq. (\ref{eq15}) can certainly be used for very small $\alpha$ with fair confidence, it fails as soon as either $\alpha$ or $\kappa$ are of order $1$. The forthcoming conclusions will not be based on approximate equations but on the consistent integration 
of the whole system in different regimes. 

\begin{figure}
\begin{center}
\begin{tabular}{|c|c|}
      \hline
      \hbox{\epsfxsize = 7.7 cm  \epsffile{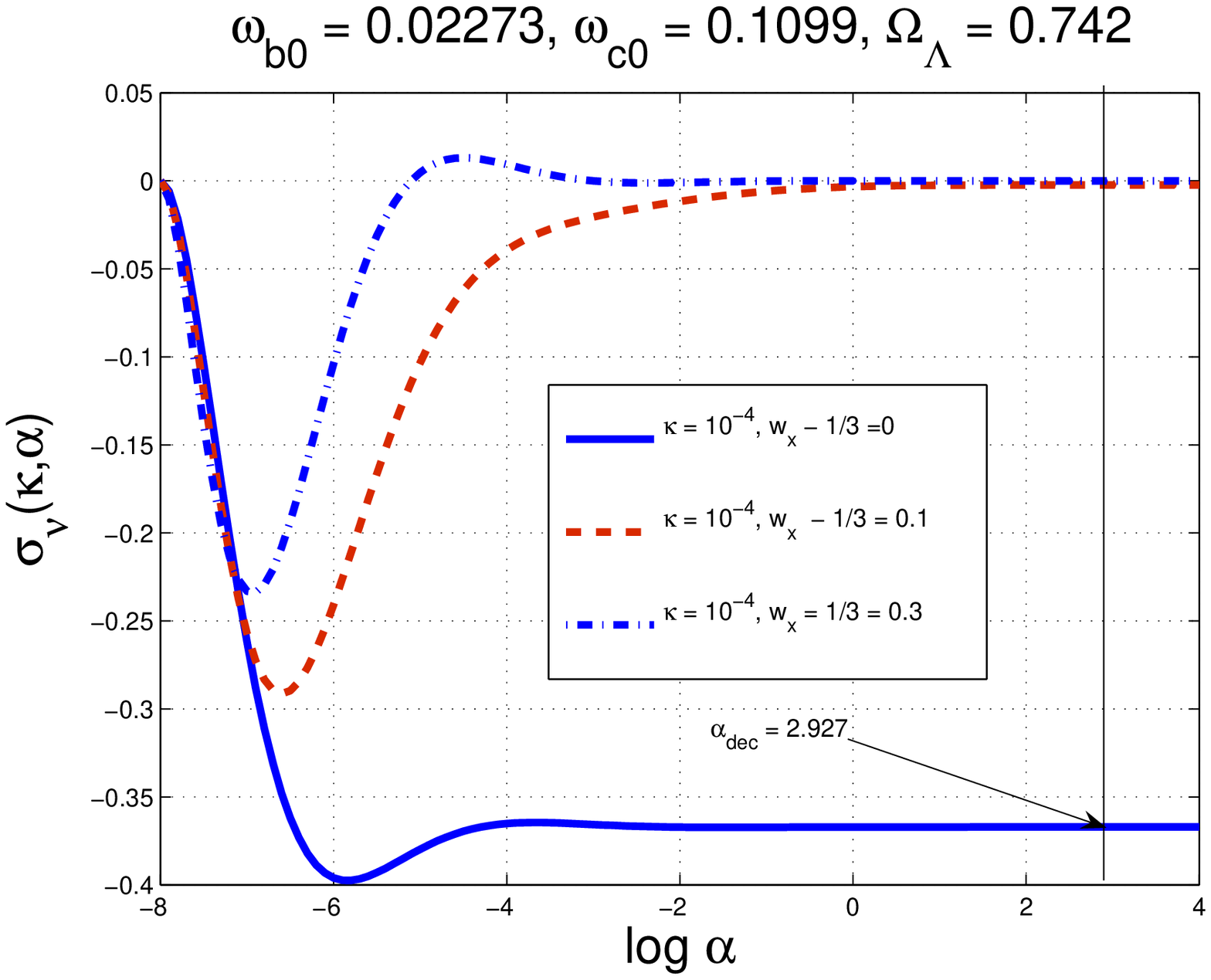}} &
      \hbox{\epsfxsize = 7.7 cm  \epsffile{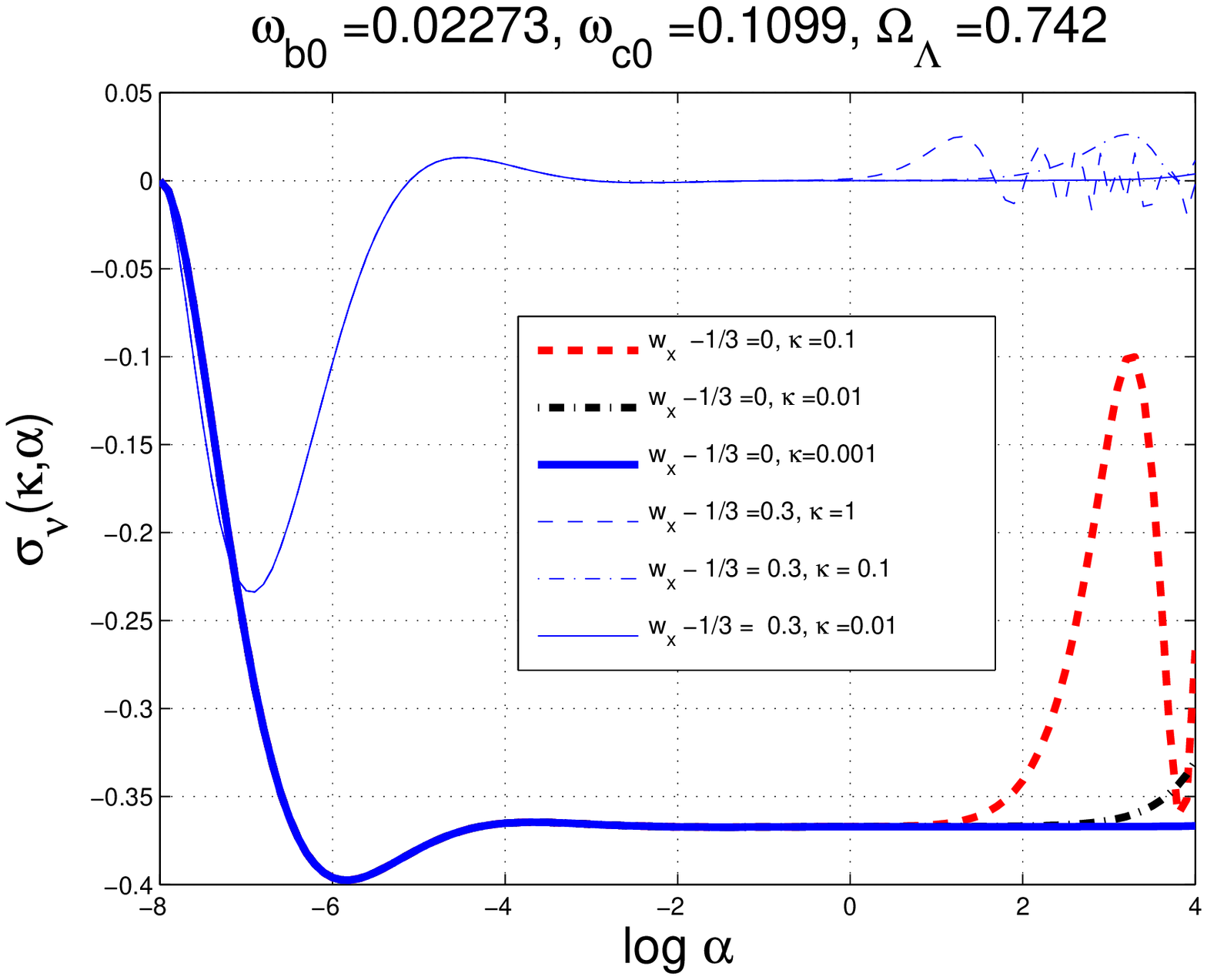}}\\
      \hline
\end{tabular}
\end{center}
\caption[a]{The relaxation of the neutrino anisotropic stress for different values of $w_{x}$ and $\kappa$.}
\label{F1}
\end{figure}

There are diverse initial conditions which could be imposed to the whole system. They are customarily divided into adiabatic and non-adiabatic \cite{hannu1,KS}(see also \cite{MB1,MB2}). In what follows we shall be interested 
in enforcing the adiabaticity condition by demanding that, at the onset of the dynamical evolution, the Hamiltonian and the momentum constraints are satisfied. The latter considerations imply that, quite generically \footnote{It should be borne in mind that, 
in the synchronous gauge, the condition $\theta_{\mathrm{c}}$ is enforced not so much because of a property of the initial 
data but rather to fix completely the coordinate system and to avoid the occurrence of known spurious (gauge) modes).}
\begin{eqnarray}
&& \delta_{\nu}(\kappa,\alpha_{\mathrm{i}}) \simeq \delta_{\gamma}(\kappa,\alpha_{\mathrm{i}}) \simeq \frac{3}{4} \delta_{\mathrm{b}}(\kappa,\alpha_{\mathrm{i}}) \simeq \frac{3}{4} \delta_{\mathrm{c}} = - R_{\gamma}, \qquad 
\sigma_{\nu}(\kappa, \alpha_{\mathrm{i}}) =0,
\Omega_{x}(\kappa,\alpha_{\mathrm{i}}),
\label{IN1}\\
&& \theta_{\nu}(\kappa,\alpha_{\mathrm{i}}) \simeq \theta_{\gamma \mathrm{b}}(\kappa,\alpha_{\mathrm{i}}) \simeq \theta_{\mathrm{c}}(\kappa,\alpha_{\mathrm{i}}) \simeq 0,\qquad \sigma_{x}(\kappa,\alpha_{\mathrm{i}}) = \frac{3 w_{x}}{4} \Omega_{x}(\kappa,\alpha_{\mathrm{i}}),
\label{IN2}
\end{eqnarray}
where $\alpha_{\mathrm{i}}$ denotes the initial integration time when the neutrino anisotropic stress vanishes exactly.
We are now interested to see what happens to the neutrino anisotropic stress when the initial conditions 
of the system obey Eqs. (\ref{IN1}) and (\ref{IN2}). The results of the numerical integration are illustrated in Fig. \ref{F1} 
for different values of $w_{x}$ and different values of $\kappa$.
The initial conditions are set in the limit $\alpha \ll \alpha_{\mathrm{dec}}$ where $\alpha_{\mathrm{dec}} = 
a_{\mathrm{dec}}/a_{\mathrm{eq}}$ denotes the value of $\alpha$ at photon decoupling (for the best 
fit to the 5yr WMAP data alone $\alpha_{\mathrm{dec}} \simeq 2.92$ as indicated in Fig. \ref{F1} with the vertical line 
in the plot at the left).  If $w_{x} > 1/3$  the anisotropic stress is driven to zero; if 
$\kappa \ll 1$, then $\sigma_{\nu}(\kappa,\alpha)$ will not oscillate for $\alpha \geq 1$ while in the opposite case (i.e. $\kappa <1$) the neutrino anisotropic stress will 
be oscillating for the same range of $\alpha$. If $w_{x}  = 1/3$ the asymptotic value of $\sigma_{\nu}(\kappa,\alpha)$ will be, approximately, $-  R_{\gamma}/R_{\nu} \sigma_{x}(k,\alpha_{\mathrm{dec}})$ for $\kappa < 1$; in the opposite case 
(i.e. $\kappa>1$)  $\sigma_{\nu}$ will oscillate around the same asymptote reached in the $\kappa < 1$ case. A similar
 phenomenon has been discussed, in the past, in the case of large-scale magnetic fields which 
  affect the anisotropic stress but which also interact with the charged particles \cite{mm}. Finally notice that,
  if $w_{x} < 1/3$, $\Omega_{x}(\kappa,\alpha)$ grows with $\alpha$ and might even get 
  dominant. This situation is more similar to the one treated in \cite{refdm} and will not 
  be specifically addressed here.  From the numerical solution 
  of the system in terms of $h(\kappa,\alpha)$ and $\xi(\kappa,\alpha)$ one can also 
  compute other fluctuations with relevant gauge-invariant interpretation such as the curvature perturbations 
  on comoving orthogonal hypersurfaces (i.e. ${\mathcal R}$), the density contrast on comoving 
  orthogonal hypersurfaces (i.e.  $\epsilon_{\mathrm{m}}$) and also $\zeta$ the curvature perturbations on constant 
  density hypersurfaces (see, e.g. \cite{JHN}) :
  \begin{eqnarray}
  && {\mathcal R} = \xi + \frac{2 \alpha (\alpha+1)}{(3 \alpha + 4)}\frac{\partial \xi}{\partial \alpha}, \qquad 
  \epsilon_{\mathrm{m}} = \frac{\kappa^2 \alpha^2}{6 (\alpha +1)} \xi - \frac{\alpha}{3} \frac{\partial Q}{\partial\alpha},
\qquad \zeta = {\mathcal R} + \frac{\alpha +1}{3\alpha + 4} \epsilon_{\mathrm{m}}.
\label{Rzetaeps}
\end{eqnarray}
Denoting with $\alpha_{\mathrm{i}}\ll 1$ the value of $\alpha$ at the onset of the numerical integration,
$\xi(\kappa,\alpha) \propto \ln{(\alpha/\alpha_{\mathrm{i}})}$ provided $\sigma_{\nu}(\kappa, \alpha_{\mathrm{i}}) \to 0$ 
since $\theta_{\nu}$ and $\theta_{\gamma\mathrm{b}}$ are both proportional to $\kappa^2 \alpha$.

\begin{figure}
\begin{center}
\begin{tabular}{|c|c|}
      \hline
      \hbox{\epsfxsize = 7.7 cm  \epsffile{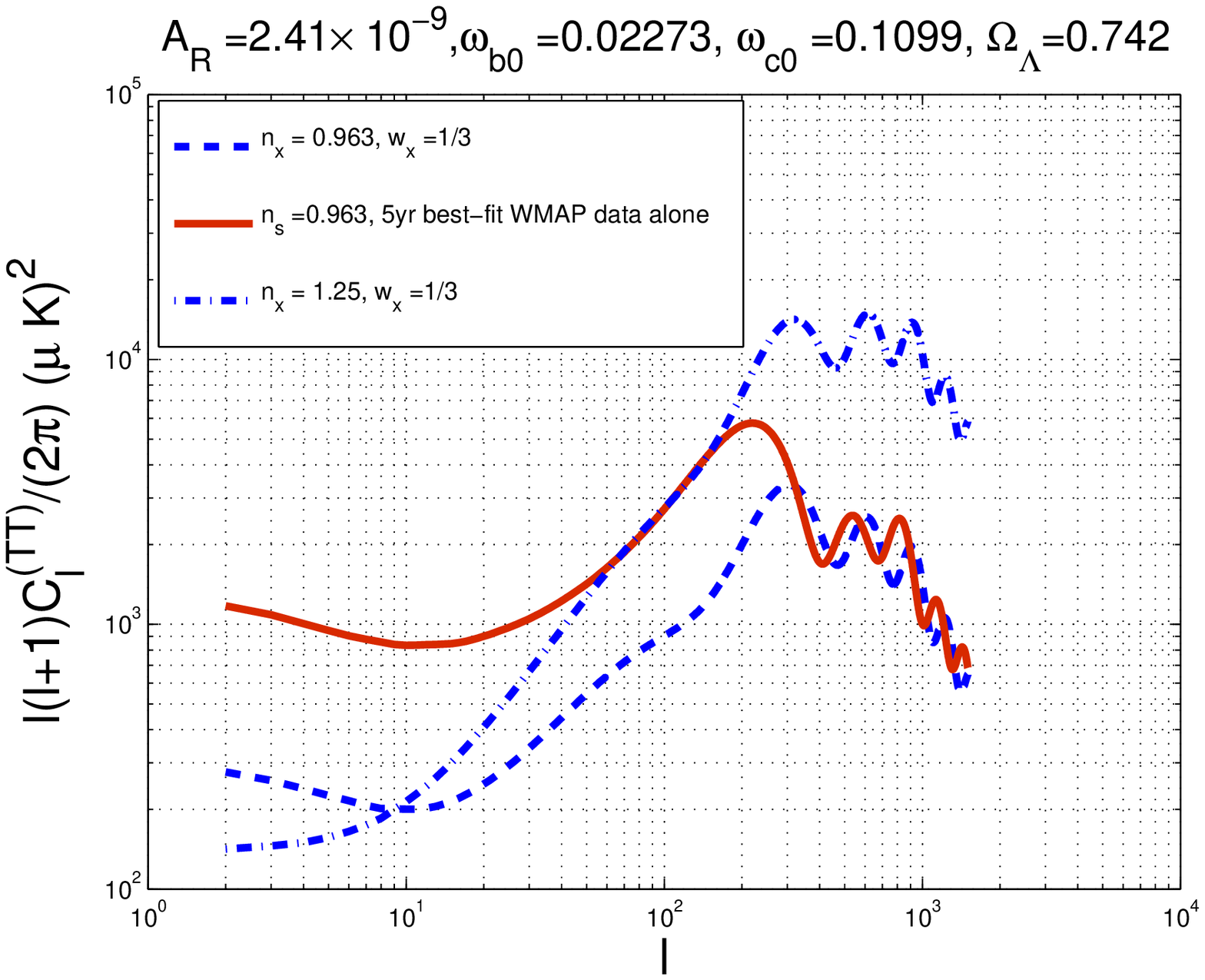}} &
      \hbox{\epsfxsize = 7.7 cm  \epsffile{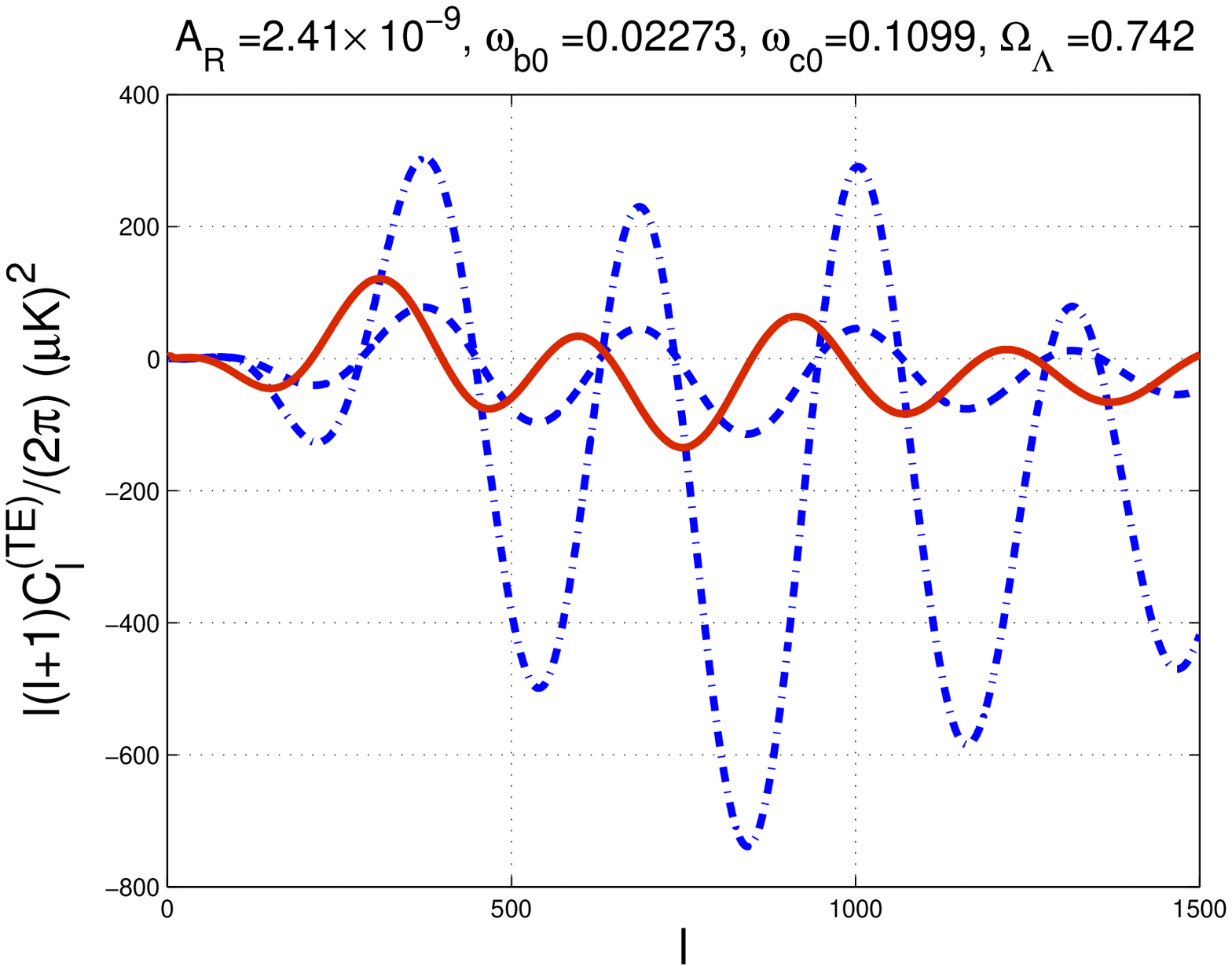}}\\
      \hline
\end{tabular}
\end{center}
\caption[a]{The TT and TE power spectra induced by the spectator anisotropic stress in the absence 
of the conventional adiabatic mode (dashed and dot-dashed lines in both plots). The WMAP 5yr best fit is reported, for 
comparison, with the full lines in both plots. Double logarithmic scale is used in the plot at the left while a linear scale 
is employed in the plot at the right.}
\label{F2}
\end{figure}

It is interesting to assume now that the whole source of large-scale inhomogeneity resides 
in the spectator anisotropic stresses. Could we get reasonable shapes of the CMB observables in the absence 
of the conventional adiabatic mode?  The answer is negative and it is contained in Fig. \ref{F2} 
where the temperature autocorrelations (i.e. TT power spectra) and the temperature-polarization 
cross-correlations (i.e. TE power spectra) are computed in the most favourable case, i.e.
$w_{x} = 1/3$. 
In both plots of Fig. \ref{F2}  the full line illustrates the WMAP 5yr best fit (in terms of the WMAP data 
alone and in the case of the conventional adiabatic mode) while the dashed and dot-dashed lines refer 
to the case of the spectator stresses with different spectral indices (one of them coinciding with the 
one of the standard adiabatic mode). It has been assumed that the power spectrum of $\sigma_{x}$ 
is assigned as ${\mathcal P}_{\sigma}(k) = {\mathcal B}(k/k_{\mathrm{p}})^{n_{x} -1}$. The amplitude has been 
left adjustable\footnote{By adjustable we mean that it has been taken to match, approximately, 
with the observed one. However, in spite of possible adjustements in the amplitude and well as in the pivot scale, the shapes 
of the TT and TE correlations cannot be made to coincide neither with the ones of the conventional adiabatic mode nor with the typical patterns of the isocurvature modes.} and, for the purposes of Fig. \ref{F2}, ${\mathcal B}= 2.41 \times 10^{-9}$ 
and $k_{\mathrm{p}} =0.002\, \mathrm{Mpc}^{-1}$. 
The numerical integration has been carried on by using, as initial conditions of the Boltzmann solver, the result 
of the numerical integration of the explicit system discussed above. The matching regime between the two regimes coincides
 with $\alpha\simeq 10^{-4}$ when the tight-coupling between baryons and photons is valid. The first 
regime of evolution dictated by the equations derived here avoids a potentially stiff problem if $\alpha$ is initially 
very small.  The Boltzmann solvers is a modified version of \cite{mg1,mg2} which, in turn, based 
 on \cite{MB1,MB2}. From Fig. \ref{F2} it appears that, as expected, the shapes and phases of the TT and TE 
correlations obtained in the case of the adiabatic mode (full lines in both plots) differ from the ones induced by the spectator stresses (dashed and dot-dashed lines in both plots of Fig. \ref{F2}). Spectator stresses can be treated and discussed in conjunction with a dominant adiabatic mode; absent 
the adiabatic mode the shapes of the TT correlations have intermediate features between the isocurvature humps 
of the CDM-radiation mode \cite{hannu1}; the situation is also different from the case of the magnetized CMB observables 
when the adiabatic mode is absent \cite{mg1} (see also \cite{mg2}).  There has been recently some confusing statement in the literature. For instance, in \cite{refwr} (first reference) it is claimed that large-scale magnetic fields provide the same shapes for the TT correlations obtainable in the case of an adiabatic mode with appropriate amplitude \footnote{
The initial conditions derived in \cite{mm} have been used in \cite{refwr} (with supplementary typos) and then subsequently criticized by a subset of the authors (see, respectively, third and second reference of \cite{refwr}).}. This statement is in sharp contrast with the present investigation. 

In summary, as anisotropic stresses can be attributed to a dominant fraction of the present energy (or matter) density 
\cite{refdm}, it is also plausible to speculate that there are stresses associated with species which are subdominant today. In a rather 
general framework, by enforcing covariant conservation, the initial conditions of the  Einstein-Boltzmann hierarchy have 
been numerically investigated using the normalized scale factor as pivot variable at early times. 
The shapes and phases of the temperature an polarization observables are drastically different from the ones induced by an adiabatic mode. Spectator anisotropic stresses lead then to an interesting kind of initial conditions which can be constrained using the available CMB data in analogy with what has been already done in the case of large-scale magnetic fields. This analysis is however beyond the scopes of the present paper.

\end{document}